\documentclass[a4paper]{jpconf}
\usepackage{graphicx}
\usepackage{color}
\usepackage{lmodern}
\usepackage[T1]{fontenc}
\usepackage{multirow,array}
\usepackage{mathtools}
\usepackage{float}
\usepackage{anysize}
\usepackage{amsmath}
\usepackage{hyperref}
\begin{document}

\title{Generalized equations of state and regular universes}
\author{F Contreras$^1$, N Cruz$^2$ y E Gonz\'alez$^2$}
\address{$^1$ Departamento de Matem\'atica y Ciencia de la Computaci\'on, Universidad de Santiago de Chile, Las Sophoras 173, Santiago, Chile}
\address{$^2$ Departamento de F\'isica, Universidad de Santiago de Chile, Avenida Ecuador 3493, Santiago, Chile}
\ead{norman.cruz@usach.cl}

\begin{abstract}
We found non singular solutions for universes filled with a fluid
which obey a Generalized Equation of State of the form
$P(\rho)=-A\rho+\gamma\rho^{\lambda}$. An emergent universe is
obtained if $A=1$ and $\lambda =1/2$. If the matter source is
reinterpret as that of a scalar matter field with some potential,
the corresponding potential is derived. For a closed universe, an
exact bounce solution is found for $A=1/3$ and the same $\lambda $.
We also explore how the composition of theses universes can be
interpreted in terms of known fluids. It is of interest to note that
accelerated solutions previously found for the late time evolution
also represent regular solutions at early times.
\end{abstract}


\section{Introduction}
\noindent
The study of Generalized Equations of State (EoS) for the main fluid component of the universe has a long time,
inspired, as the best of our knowledge, in the particular behavior of Friedmann models
during inflationary scenarios.  In order to extend the range of known inflationary behaviors, Barrow
~\cite{Barrow} assumed the matter stress has pressure $P$ and density $\rho$, related by the following model equation of state
\begin{equation}
P(\rho)=-\rho+\gamma\rho^{\lambda},
\label{EoS}
\end{equation}
where $\gamma$ and $\lambda$ are both constants and $\gamma\neq 0$. The standard EoS of a perfect fluid, $P=(\gamma -1)\rho$
is recovered when $\lambda =1$. An emergent flat universe solution was found for the case $\lambda =1/2$ and $\gamma >0$ although
not discussed in this context. In this solution $a(t\rightarrow \infty)\rightarrow \infty$ and $a(t\rightarrow -\infty)\rightarrow 1$.
The Hubble parameters and its derivatives were not investigated in detail. It is interesting to mention that the doubled
exponential behavior of this solution was previously found for a bulk viscous source in the presence of an effective
cosmological constant~\cite{Barrow1}. This is a consequence of the inclusion of bulk viscosity in the Eckart's theory, which leads to and viscous
pressure $\Pi$ of the form $-3\xi H$, where $\xi$ is assumed usually in the form $\xi=\xi_{0} \rho^{\lambda}$.

A variation of Eq.(\ref{EoS}) was considered by Mukherjee \textit{et al}~\cite{Mukherjee} assuming the form
\begin{equation}
P(\rho)=-A\rho-\gamma\rho^{1/2},
\label{EoS1}
\end{equation}
where for $A<1$ and $\gamma >0$ it is possible to obtain a solution describing an emergent flat universe, by with a scale factor
of the form
\begin{equation}
a(t)=a_{0}(\beta +e^{\alpha t})^{\omega},
\label{aemergent}
\end{equation}
where $a_{0}$ and $\beta$ are constants and $\alpha$, $\omega$ are
given in terms of the parameters $A,\gamma$. For same values of $A$
the energy density of this fluid can be interpreted as the sum of
three known fluids such radiation, string and a cosmological
constant. The case with $A=1$ and $\lambda=1/2$ was considered in
~\cite{Odintsov}, ~\cite{Stefancic}. In both works the cosmological
solutions of dark energy models with this fluid was analyzed,
focusing in the future expansion of the universe. A late time
behavior of a universe filled with a dark energy component with an
EoS given by Eq.(\ref{EoS}) has been investigated in ~\cite{Paul},
~\cite{Paul1}, where the allowed values of the parameters $A$ and
$\gamma$ were constrained using H(z)-z data, a model independent BAO
peak parameter an cosmic parameter (WMAP7 data).

The research of late time evolution of the universe has also assumed
EoS of the type given by Eq.(\ref{EoS}) motivated by the fact that
the constraints from the observational data implies $\omega \approx
-1$ for the EoS of the dark energy component if it assumed ruled by
a barotropic EoS, but the values $\omega < -1$, corresponding to a
phantom fluid, or $\omega > -1$, corresponding to quintessence can
not be discarded. Also theoretical studies like the so called
running vacuum energy in QFT (see ~\cite{Shapiro}) gives rise to a
cosmological constant with a dynamical evolution during the cosmic
time, so EoS of the type given in Eq.(\ref{EoS}) could represent
effectively the results of this approach under some specific
assumptions.

The aim of the present work is to explore non singular solutions for
universes filled with a fluid which obey and EoS given by
Eq.(\ref{EoS}). In particular, in the next section we show that
previous solutions found for a late time solutions represents also
an emergent universe. We evaluate also the field potential $V(\phi)$
if the matter source is reinterpreted as that of a scalar matter.

In section 3 we found an exact solution for a bouncing universe with positive curvature. We show that for this case the behavior of the density energy and pressure is like that tree known fluids. In section 4 we discuss our results.


\section{Exact solution for an emergent universe}
\noindent
As it was mentioned above an EoS of the form given in Eq.(\ref{EoS}) leads to an emergent flat universe solution for the case $\lambda =1/2$, i. e., for an EOS given by
\begin{equation}
P(\rho)=-\rho-\gamma\rho^{1/2}.
\label{EoSA1}
\end{equation}

The Einstein equations for a flat FRW metric without cosmological constant can be write in the following form
\begin{equation}
\rho=3\left( \frac{\dot{a}}{a}\right) ^{2},
\label{FRW1}
\end{equation}
\begin{equation}
P=-2\left( \frac{\ddot{a}}{a}\right)-\left( \frac{\dot{a}}{a}\right)^{2}.
\label{FRW2}
\end{equation}
From the equations (\ref{EoSA1}) to (\ref{FRW2}) and assuming $\dot{a},a \neq 0$ , we obtain
\begin{equation}
\frac{\dot{a}}{a}=K \exp \left( \frac{\gamma\sqrt{3}}{2}t\right),
\label{apunto}
\end{equation}
so the solution for the scale factor as a function of the cosmic time is given by
\begin{equation}
a(t)=\sigma \exp \left[\frac{2K}{\gamma\sqrt{3}}\exp\left( \frac{\gamma\sqrt{3}}{2}t\right)\right],
\label{adet}
\end{equation}
where $K$ y $\sigma$ are positive integration constants. This is the
solution already found by Barrow~\cite{Barrow}.  Note that for
$\gamma<0$, $a (t \rightarrow -\infty) \rightarrow 0$ and we have
zero scale factor in the infinity past. An emergent universe
solution is obtained for $\gamma>0$, since $a (t \rightarrow
-\infty) \rightarrow \sigma$ so the scale factor tends to a finite
value different from zero in the infinity past. If we choose the
initial conditions $a(t_{0})=a_{0}$ and $\rho(t_{0})=\rho_{0}$ the
integration constants $K$ y $\sigma$ are given by
\begin{equation}
K=\frac{\rho_{0}^{1/2}}{\sqrt{3}} \exp \left( -\frac{\gamma\sqrt{3}}{2}t_{0}\right), \,\,\, \sigma =a_{0}\exp \left( -\frac{2\rho_{0}^{1/2}}{3\gamma}\right).
\label{Kysigma}
\end{equation}
Then the solution for the scale factor with these initial conditions becomes
\begin{equation}
a(t)=a_{0} \exp \left( \frac{2\rho_{0}^{1/2}}{3\gamma}\left[\exp \left( \frac{\gamma\sqrt{3}}{2}(t-t_{0})\right)-1 \right] \right),
\label{adet1}
\end{equation}
and the Hubble parameter and its derivatives are given by
\begin{equation}
H(t)=\frac{\rho_{0}^{1/2}}{\sqrt{3}}\exp\left[\frac{\gamma\sqrt{3}}{2}(t-t_0)\right],
\label{H}
\end{equation}
\begin{equation}
\dot{H}(t)=\frac{\rho_{0}^{1/2}\gamma}{2}\exp\left[\frac{\gamma\sqrt{3}}{2}(t-t_0)\right],
\label{Hdet}
\end{equation}
\begin{equation}
\ddot{H}(t)=\frac{\rho_{0}^{1/2}\gamma^{2}\sqrt{3}}{4}\exp\left[\frac{\gamma\sqrt{3}}{2}(t-t_0)\right].
\label{Hdospuntos}
\end{equation}
The above expressions indicates that $H$, $\dot{H}$ and $\ddot{H}$
are positive. Thus $H$, $\dot{H}$ and $\ddot{H}$ tends to zero for
$t\rightarrow -\infty$ and to infinity when $t\rightarrow \infty$.

\begin{figure}[H]
\begin{center}
\includegraphics[width=20pc]{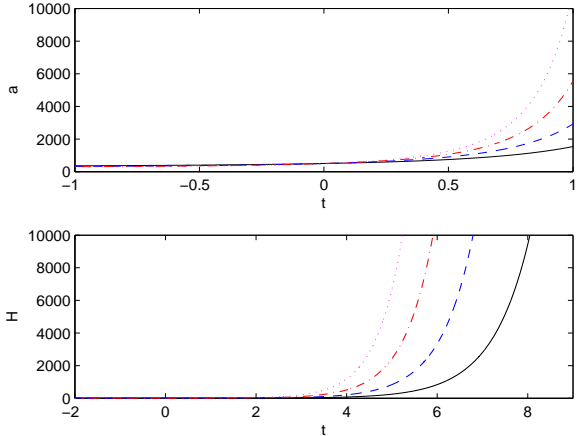}\hspace{2pc}
\begin{minipage}[b]{13pc}\caption{\label{FEPH_UE}Graphics scale factor (upper) and the Hubble parameter (lower) as a function of time for the emergent universe with $a_0=500$ and $t_0=0$. The black line ($\full$) is for $\gamma=1.4$ and $\rho_0=1$, blue line ($\longbroken$) for $\gamma=1.6$ and $\rho_0=2$, red line ($\chain$) for $\gamma=1.8$ and $\rho_0=3$ and magenta line ($\dotted$) for $\gamma=2.0$ and $\rho_0=4$.}
\end{minipage}
\end{center}
\end{figure}


\subsection{Composition of the emergent universe}
\noindent
In the following we shall study the energy density as a function of the scale factor
in order to looking for a possible equivalence between our fluid in terms of other knows fluids. Using the energy conservation equation
\begin{equation}
\dot{\rho}+3\frac{\dot{a}}{a}(\rho+P)=0,
\label{conservacion}
\end{equation}
and taken the EoS given by Eq.(\ref{EoSA1}) we obtain
\begin{equation}
\rho(a)=\left[ \frac{3\gamma}{2}\ln \left(\frac{a}{a_{0}} \right)+\rho_{0}^{1/2}\right]^{2}.
\label{rhodea}
\end{equation}
Using Eq. (\ref{adet1}) in Eq. (\ref{rhodea}) we obtain the energy as a function of time
\begin{equation}
\rho(t)=\rho_{0}\exp \left( \gamma\sqrt{3}(t-t_{0})\right),
\label{rhodet}
\end{equation}
which indicates that $\rho$ tends to zero for $t\rightarrow$ $-\infty$. Introducing Eq.(\ref{rhodea})
in Eq.(\ref{EoSA1}) we obtain the fluid pressure as a function of the scale factor, which is given by
\begin{equation}
P(a)=- \left[ \frac{3\gamma}{2}\ln \left(\frac{a}{a_{0}} \right)+\rho_{0}^{1/2}\right]^{2}-\gamma \left[ \frac{3\gamma}{2}\ln \left(\frac{a}{a_{0}} \right)+\rho_{0}^{1/2}\right].
\label{pdea}
\end{equation}
In order to obtain now the fluid pressure as a function of time we use Eq.(\ref{rhodet}) in Eq.(\ref{EoSA1})
\begin{equation}
P(t)=- \rho_{0}\exp \left( \gamma\sqrt{3}(t-t_{0})\right)- \gamma \rho_{0}^{1/2}\exp \left( \frac{\gamma\sqrt{3}}{2}(t-t_{0})\right).
\label{pdet}
\end{equation}

\begin{figure}[h]
\begin{center}
\begin{minipage}{17pc}
\includegraphics[width=17pc]{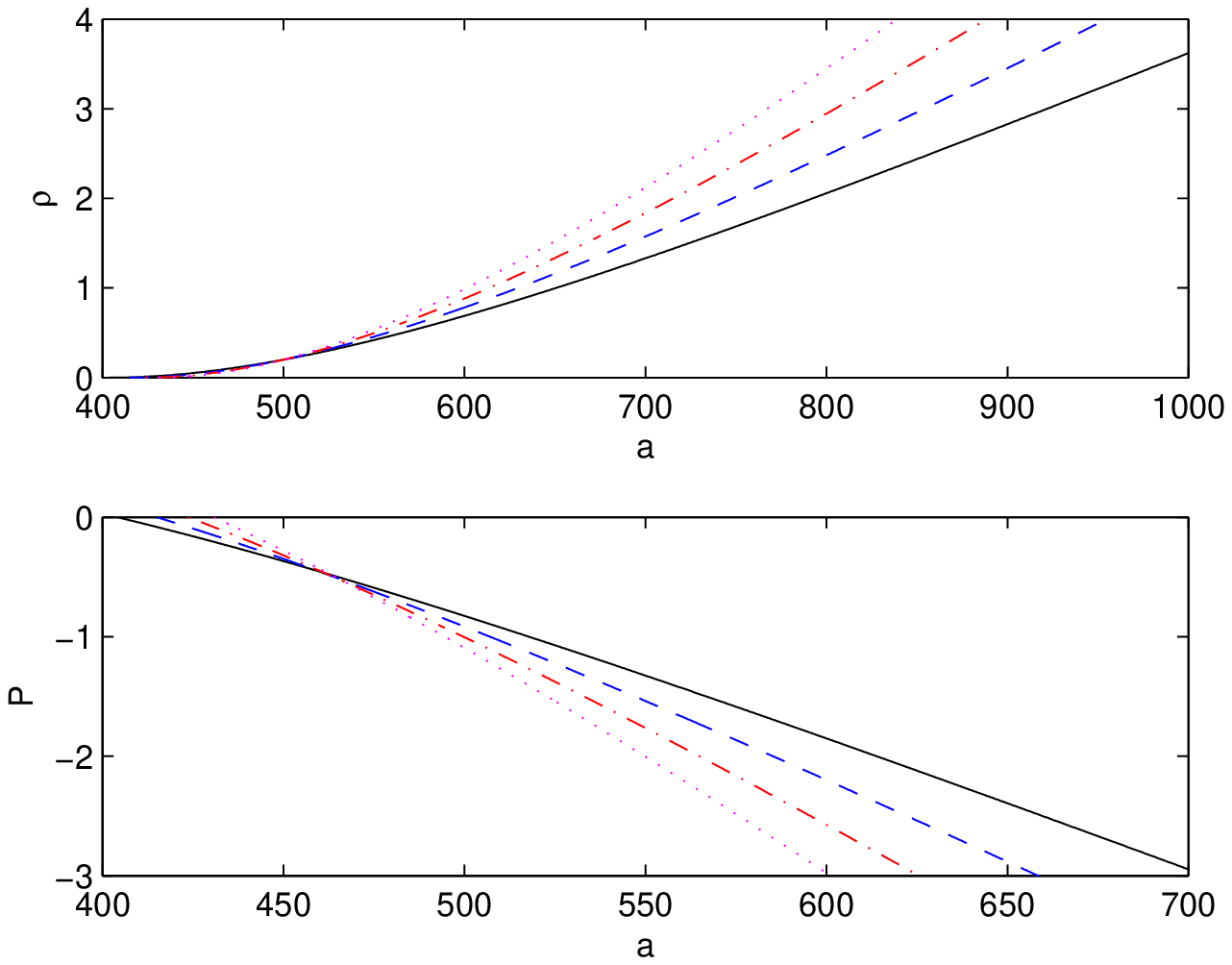}
\caption{\label{DEYP_UE}Graphic energy density (upper) and pressure (lower) as a function of scale factor for the emergent universe with $a_0=500$, $t_0=0$ and $\rho_0=0.2$. The black line ($\full$) is for $\gamma=1.4$, blue line ($\longbroken$) for $\gamma=1.6$, red line ($\chain$) for $\gamma=1.8$ and magenta line ($\dotted$) for $\gamma=2.0$.}
\end{minipage}\hspace{2pc}
\begin{minipage}{17pc}
\includegraphics[width=17pc]{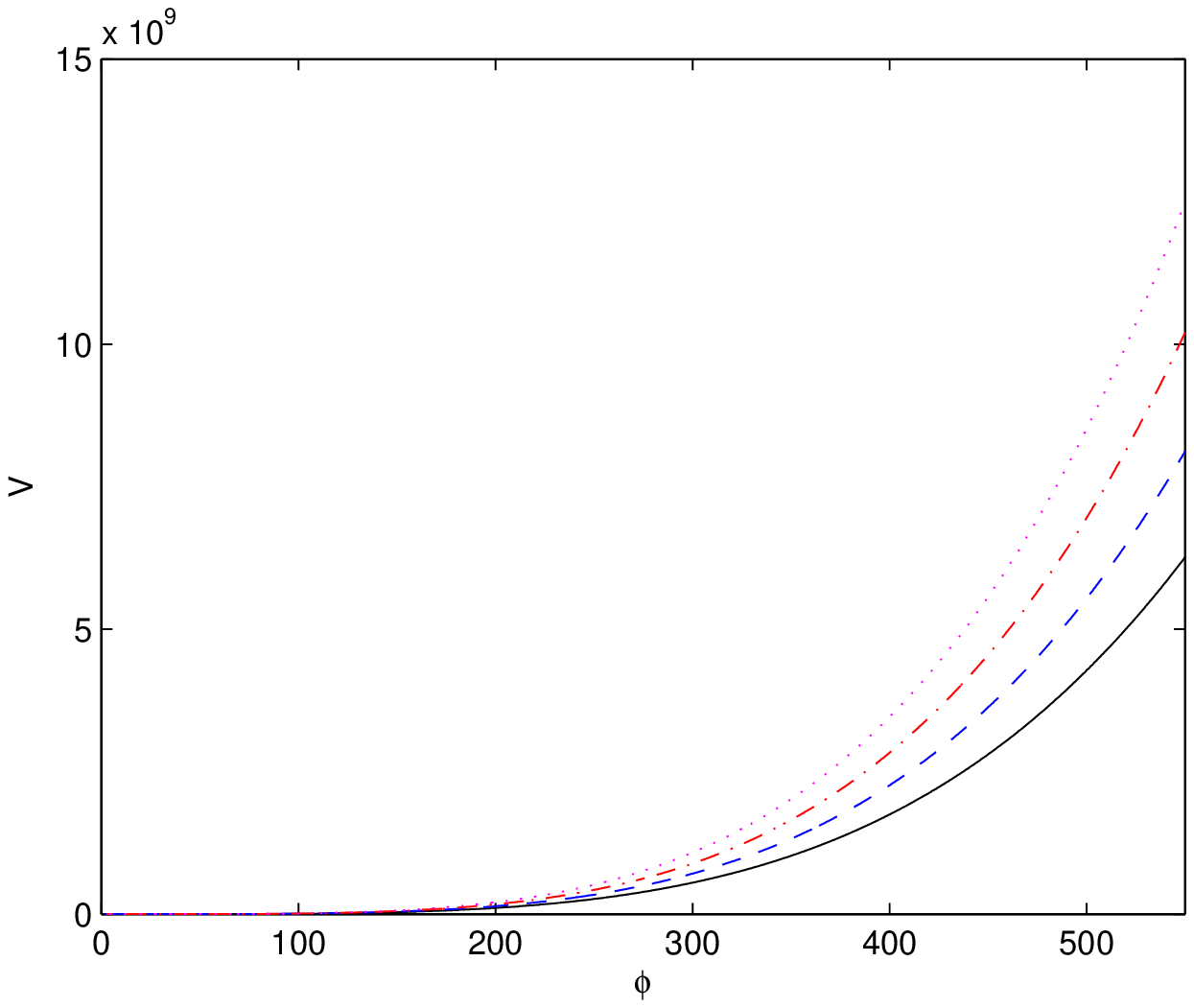}
\caption{\label{P_UE}Graphic of the scalar field potential for the emergent universe with $t_0=0$. The black line ($\full$) is for $\gamma=1.4$, $\phi_0=1$ and $\rho_0=0.2$, blue line ($\longbroken$) for $\gamma=1.6$, $\phi_0=2$ and $\rho_0=0.4$, red line ($\chain$) for $\gamma=1.8$, $\phi_0=3$ and $\rho_0=0.6$ and magenta line ($\dotted$) for $\gamma=2.0$, $\phi_0=4$ and $\rho_0=0.8$.}
\end{minipage}
\end{center}
\end{figure}

For the EoS given by Eq.(\ref{EoSA1}) we can reinterpret the matter
source as that of a scalar matter field with some potential
$V(\phi)$. Since a $\omega = P/\rho$, for the EoS considered we
obtain
\begin{equation}
\omega=-1-\frac{\gamma}{\rho^{1/2}},
\label{omegaphantom}
\end{equation}
so we always have a phantom behavior for $\gamma >0$. Therefore, Friedmann equations can be solved taken in the energy density and in the pressure a negative kinetic term
\begin{equation}
\rho=-\frac{1}{2}\dot{\phi}^{2}+V(\phi),
\label{rhofield}
\end{equation}
\begin{equation}
P=-\frac{1}{2}\dot{\phi}^{2}-V(\phi).
\label{pfield}
\end{equation}
Using Eqs. (\ref{EoSA1}), (\ref{FRW1}) and (\ref{rhofield}), (\ref{pfield}) yields
\begin{equation}
V(\phi)=\frac{3\gamma^{2}}{256}[3(\phi-\phi_0)^{4}+8(\phi-\phi_0)^{2}],
\label{Vdephi}
\end{equation}
for the field potential. In~\cite{Barrow1} a similar but negative expression was found for
the potential, but without the inclusion of phantom matter. The behavior of field $\phi$
and the potential as a function of time is easily obtained
\begin{equation}
\phi(t)=\frac{4\rho_{0}^{1/4}}{\sqrt{3\gamma}}\exp \left( \frac{\gamma\sqrt{3}}{4}(t-t_{0})\right)+ \phi_{0},
\label{phidet}
\end{equation}
\begin{equation}
V(t)=\rho_{0}\exp \left( \gamma\sqrt{3}(t-t_{0})\right)+ \frac{\gamma\rho_{0}^{1/2}}{2}\exp \left( \frac{\gamma\sqrt{3}}{2}(t-t_{0})\right).
\label{Vdet}
\end{equation}
Since the energy density decrease as $t$ goes to $-\infty$, the $\omega$ parameter becomes smaller than $-1$ and $\omega(t \rightarrow -\infty)=-\infty$ (see the upper plot in Fig. \ref{W_UEUB}). The above emergent solution was found in the context of phantom cosmologies without big rip singularity, so discussed for late times ~\cite{Odintsov1}


\section{Exact solution for a bouncing universe}
\noindent
For a closed universe exist a family of bouncing solutions if the parameter $A$ of Eq.(\ref{EoS1}) is taken close to the value $1/3$. These solutions can be obtained solving numerically the equation of motion.

For a universe with positive curvature ($k=1$), Eq. (\ref{FRW1}) becomes
\begin{equation}
\rho=3\left( \frac{\dot{a}}{a}\right) ^{2}+\frac{3}{a^{2}},
\end{equation}
and taken $A=1/3$ in Eq.(\ref{EoS1}) yields
\begin{equation}
P(\rho)=-\frac{1}{3}\rho-\gamma\rho^{1/2}.
\label{EoS2}
\end{equation}
In this case the following exact solution is found
\begin{equation}
a(t)=\frac{2}{\gamma\sqrt{3}}\cosh\left( \frac{\gamma\sqrt{3}}{2}t+\alpha\right)+\beta,
\label{abouncing}
\end{equation}
where $\alpha$ y $\beta$ are integration constants. The bouncing solution is obtained when $\gamma>0$.
This solution represents a universe expanding exponentially for $t=-\infty,\infty$. The scale factor takes a minimum value $a(t=t_{min})= \frac{2}{\gamma \sqrt{3}}+ \beta$, where $t_{min}=-\frac{2\alpha}{\gamma \sqrt{3}}$. The positivity of the scale factor constraints $\beta$ to be in the following range $(-\frac{2}{\gamma \sqrt{3}},\infty)$.

\begin{figure}[H]
\begin{center}
\includegraphics[width=20pc]{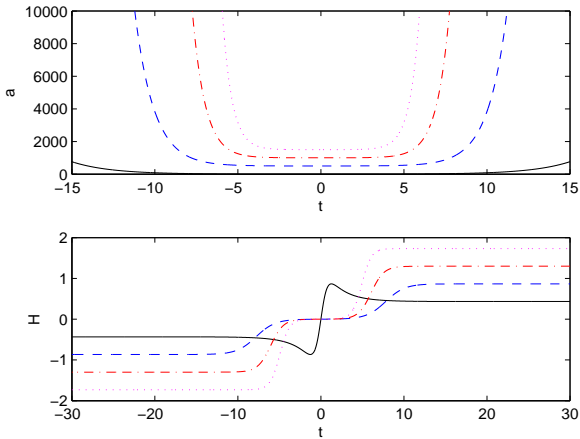}\hspace{2pc}
\begin{minipage}[b]{13pc}\caption{\label{FEPH_UB}Graphics scale factor (upper) and the Hubble parameter (lower) as a function of time for the bouncing universe with $\alpha=0$. The black line ($\full$) is for $\gamma=0.5$ and $\beta=-2$, blue line ($\longbroken$) for $\gamma=1.0$ and $\beta=500$, red line ($\chain$) for $\gamma=1.5$ and $\beta=1000$ and magenta line ($\dotted$) for $\gamma=2.0$ and $\beta=1500$.}
\end{minipage}
\end{center}
\end{figure}

Using the above solution we can evaluate $H$, $\dot{H}$ and $\ddot{H}$, which are given by the following expressions
\begin{equation}
H(x)=\frac{\sinh(x)}{\frac{2}{\gamma\sqrt{3}}\cosh(x)+\beta},
\end{equation}
\begin{equation}
\dot{H}(x)=\frac{1+\frac{\gamma\sqrt{3}}{2}\beta \cosh{ \left(x\right)}}{ \left[ \frac{2}{\gamma\sqrt{3}}\cosh { \left(x\right)}+\beta \right]^{2}},
\end{equation}
\begin{equation}
\ddot{H}(x)=\frac{\sinh{(x)}\left(\frac{3\gamma^{2}\beta^{2}}{2}-4\right)-\frac{\gamma\sqrt{3}\beta}{2}\sinh{(2x)}}{2\left[ \frac{2}{\gamma\sqrt{3}}\cosh{(x)}+\beta \right]^{3}},
\end{equation}
where $x=\frac{\gamma\sqrt{3}}{2}t+\alpha$. For $\beta >0$ the Hubble parameter is a strictly increasing function, so there are no critical points and we have $ H(t\rightarrow -\infty)= -\frac{\gamma\sqrt{3}}{2}$ and $ H(t\rightarrow \infty)= \frac{\gamma\sqrt{3}}{2}$. So for the late times this solution behaves as a de Sitter phase.


\subsection{Composition of the bouncing universe}
\noindent
We can obtain the energy density as a function of the scale factor by introducing the EoS given by Eq.(\ref{EoS2}) in the continuity equation  (\ref{conservacion})
\begin{equation}
\rho(a)=\left(\frac{3\gamma}{2}+\frac{\delta}{2a}\right)^{2},
\label{rhodeaEoS2}
\end{equation}
where $\delta =-3\gamma \beta $. Note that if $\beta =0$ then $\rho $ is constant.

Introducing Eq.(\ref{rhodeaEoS2})in Eq.(\ref{EoS2}) we obtain the fluid pressure as a function of the scale factor
\begin{equation}
P(a)=-\frac{1}{3}\left( \frac{3\gamma}{2}+\frac{\delta}{2a}\right)^{2}-\gamma\left( \frac{3\gamma}{2}-\frac{\delta}{2a}\right).
\label{pdeaEoS2}
\end{equation}

\begin{figure}[h]
\begin{center}
\begin{minipage}{17pc}
\includegraphics[width=17pc]{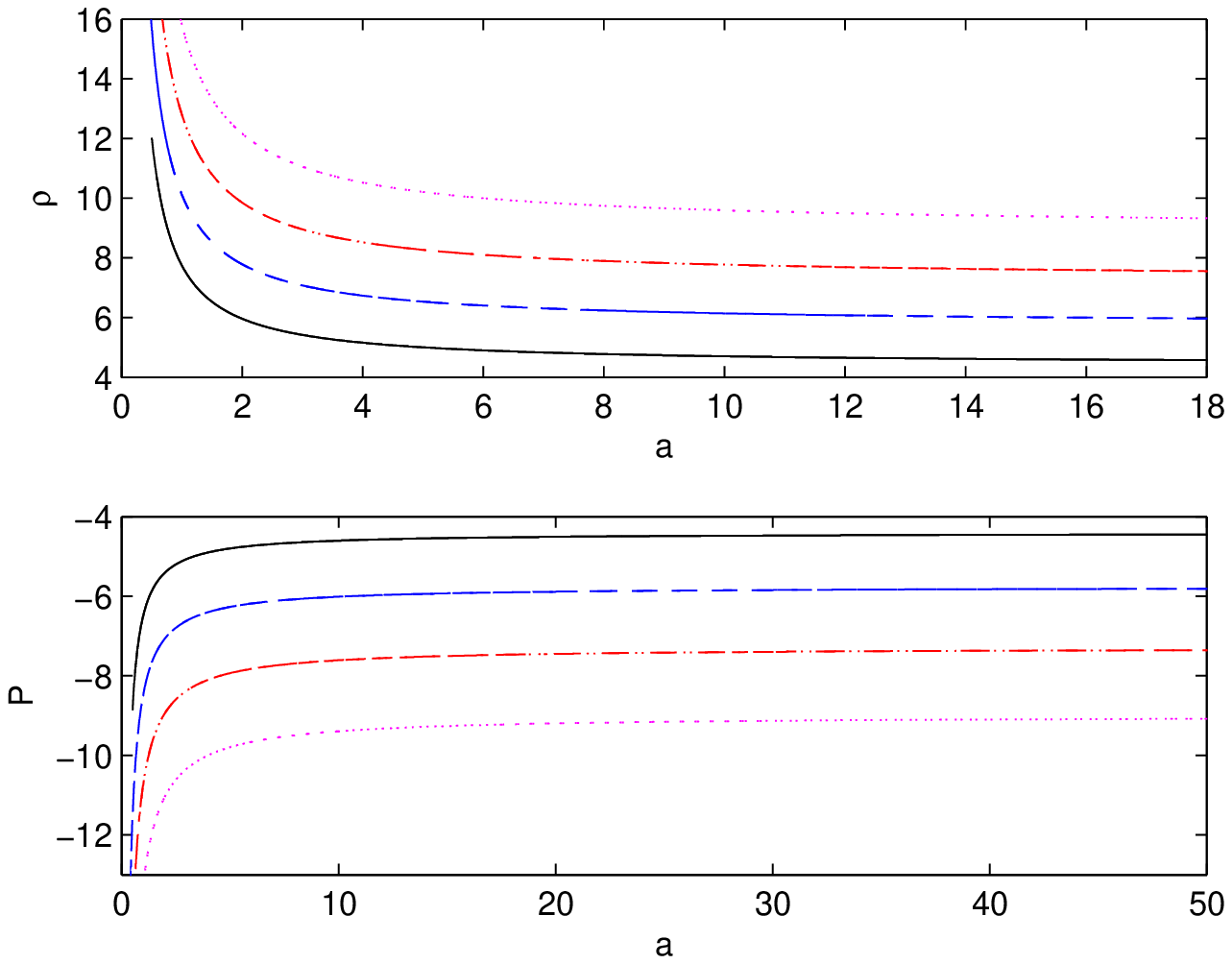}
\caption{\label{DEYP_UB}Graphic energy density (upper) and pressure (lower) as a function of scale factor for the bouncing universe with $\alpha=0$ and $\beta=-0.33$. The black line ($\full$) is for $\gamma=1.4$, blue line ($\longbroken$) for $\gamma=1.6$, red line ($\chain$) for $\gamma=1.8$ and magenta line ($\dotted$) for $\gamma=2.0$.}
\end{minipage}\hspace{2pc}
\begin{minipage}{17pc}
\includegraphics[width=17pc]{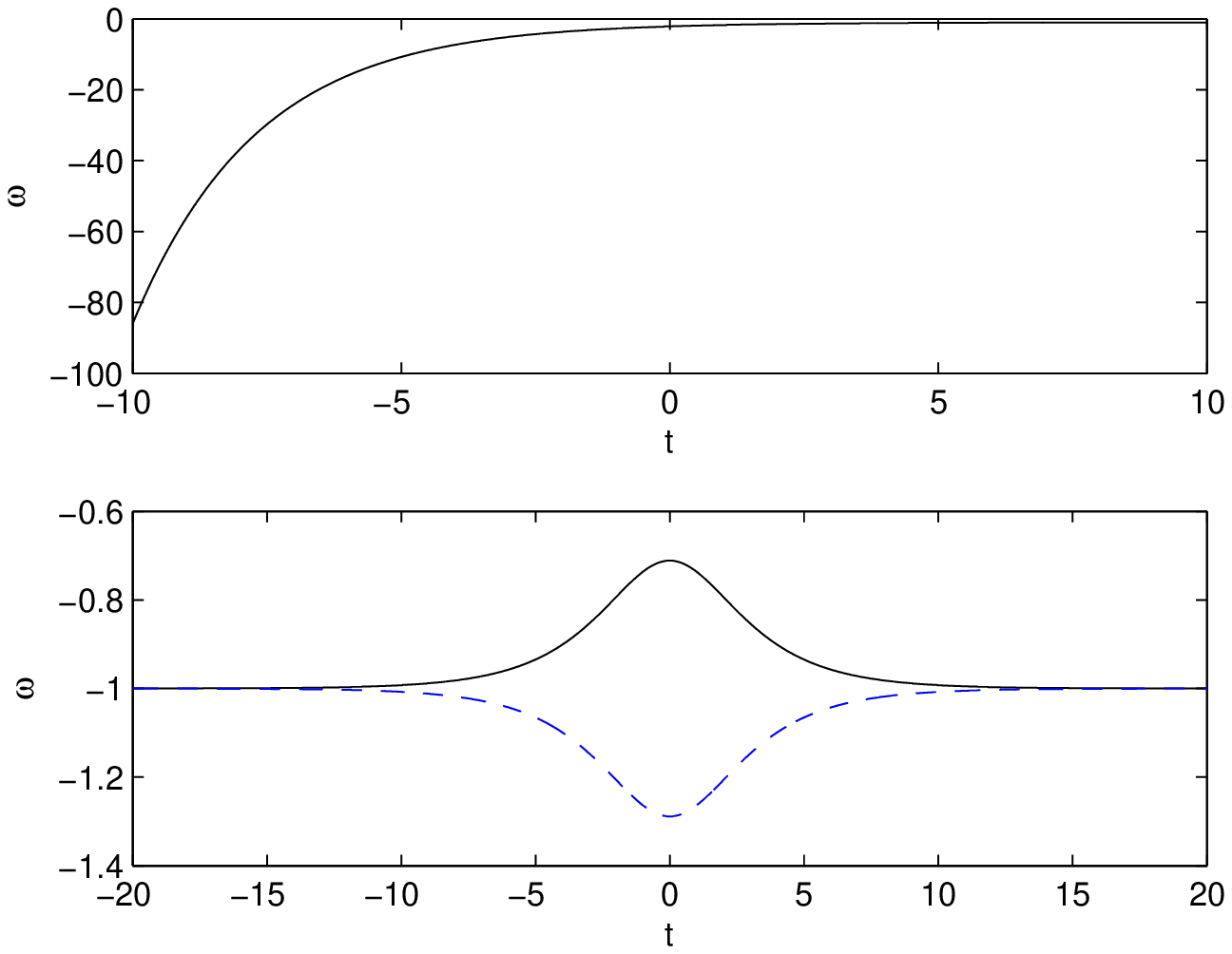}
\caption{\label{W_UEUB}Graphic of the $\omega$ parameter for emergent universe (upper) with $\gamma=0.5$, $\rho_0=0.2$, $a_0=500$ and $t_0=0$ and bouncing universe (lower) with $\gamma=0.5$ and $\alpha=0$ as a function of time. In the case of the bouncing universe, the black line ($\full$) is for $\beta=-1$ and blue line ($\longbroken$) for $\beta=1$.}
\end{minipage}
\end{center}
\end{figure}

Expanding the terms of the above both expressions yields
\begin{equation}
\rho(a)=\frac{9\gamma^{2}}{4}+\frac{3\gamma\delta}{2a}+\frac{\delta^{2}}{4a^{2}}=\rho_1+\rho_2+\rho_3,
\end{equation}
\begin{equation}
P(a)=-\frac{9\gamma^{2}}{4}-\frac{\gamma\delta}{a}-\frac{\delta^{2}}{12a^{2}}=P_1+P_2+P_3.
\end{equation}

Comparing each terms of the expansions our fluid can be seen as the sum of three fluids with the EoS given by $\omega_{1}= P_{1}/\rho_{1}=-1$, $\omega_{2}= P_{2}/\rho_{2}=-2/3$ and $\omega_{3}= P_{3}/\rho_{3}=-1/3$, respectively.  So the first fluid correspond to a cosmological constant, the second is a quintessence and the third correspond to the presence of positive curvature. In this case the used EoS has a $\omega = P/\rho$ given by
\begin{equation}
\omega=-\frac{1}{3}-\frac{\gamma}{\rho^{1/2}}.
\label{omegabouncing}
\end{equation}
Introducing Eq.(\ref{abouncing}) in Eq.(\ref{rhodeaEoS2}) we can obtain an expression for the energy density as a function of the cosmic time. Using this in Eq.(\ref{omegaphantom}) the behavior for the $\omega$ parameter can be obtained as a function as time. The lower plot in Fig. \ref{W_UEUB} depicted this behavior. It can seen that in the case with $\beta=-1$ the fluid ruled by the EoS given in Eq.(\ref{omegabouncing}) behaves like a cosmological constant for $t \rightarrow +\infty$ and $t \rightarrow -\infty$ and like quintessence for the lapse associated at the time of bouncing.


\section{Discussion}
\noindent
We have show that cosmological solutions for a flat universe filled with a generalized EoS which represent accelerated expansion for late time evolution,
also represent emergent universes for $t \rightarrow -\infty$. The behavior of the Hubble parameter indicates that these solutions are geodesically
complete. For a scalar matter field with the EoS given by Eq.(\ref{EoSA1}), the corresponding potential takes the form $V(\phi)\sim \alpha\phi^{4}+\beta\phi^{2}$.

For a closed universe and taken $A=1/3$ an exact solution was found
and the energy density and pressure evaluated as a function of the
scale factor. These expressions allows us to see that the
generalized EoS behaves in this case as the sum of a cosmological
constant, quintessence and the equivalent fluid due to curvature.
For the emergent solution the fluid always violated the NEC
condition, except for $t \rightarrow +\infty$ when the evolution
ends in a de sitter phase.

In the bouncing solution, we have the case where the NEC condition
is satisfied during the entire cosmic evolution (case $\beta<0$),
which is possible due the presence of positive curvature.


\section*{Acknowledgements}
\noindent
This work was supported by CONICYT through Grant FONDECYT N$^\circ$ 1140238 (NC), and by
Proyecto Basal, Hacia una cultura de indicadores en la Educaci\'on Superior, USA 1298 (EG).


\end{document}